\documentclass[aip,
jmp,
 amsmath,amssymb,
preprint,
]{revtex4-1}

\usepackage{graphicx}%
\usepackage{dcolumn}%
\usepackage{bm}%
\usepackage[utf8]{inputenc}
\usepackage[T1]{fontenc}
\usepackage{mathptmx}
\usepackage{etoolbox}

\makeatletter
\def\@email#1#2{%
 \endgroup
 \patchcmd{\titleblock@produce}
  {\frontmatter@RRAPformat}
  {\frontmatter@RRAPformat{\produce@RRAP{*#1\href{mailto:#2}{#2}}}\frontmatter@RRAPformat}
  {}{}
}%
\def \hg {$^{199}$Hg\ }
\makeatother

\begin{document}

\preprint{}

\title{Comment on ``Surface nuclear spin relaxation of $^{199}$Hg," [J. Chem. Phys. 120, 1511 (2004)]}

\author{S.K. Lamoreaux}

\email{steve.lamoreaux@yale.edu}
 \noaffiliation

\date{\today}

\begin{abstract}

\end{abstract}

\affiliation{Yale University, Department of Physics
Box 208120, New Haven, CT 06520-8120, USA }%

\maketitle

The use of optically pumped \hg nuclear spin magnetometers for fundamental physics studies remains of great interest.  Understanding surface-induced relaxation is necessary for the development of both large- and small-volume cells that have a long and stable nuclear spin lifetime. These cells are critical to several ongoing experiments, including several to measure a neutron electric dipole moment.\cite{lanledm,psiedm}  A previous study, with its carefully acquired large data set, \cite{romalis} has been quite helpful for understanding the dynamics of \hg spin relaxation.  The purpose of this note is to address some complications in the interpretation of these data as presented in [~\citenum{romalis}]. 

The first and foremost in with regard to Eq. (3) of [~\citenum{romalis}]. This equation is true if there is a single bound state for an \hg atom on the cell surface.  However, the number of bound surface states grows with mass,\cite{jmp} a point that has rarely, if ever, been taken into account in previous similar work (see, e.g., [\citenum{happer}]). 

The number of states and their energies can be estimated using the WKB approximation, assuming the van der Waals potential (short-range Casimir-Polder force) between an \hg atom and a dielectric surface as
\begin{equation}\label{vdw}
V(z)\approx\begin{cases} \frac{C}{(d_0 +z)^3}& z\geq 0\\
\infty & z<0 \end{cases}
\end{equation}
where $C\approx -6\times 10^{-4} \ e$V nm$^3$ for a heavy polarizable atom near a dielectric surface, with $z$ the distance from the surface in nm and $d_0\approx 0.14$ nm representing a sum of radii of the Hg atom and atoms bound in the fused silica surface.  The coefficient can be calculated using the formalism in [~\citenum{vdw95}] with the dielectric function for fused SiO$_2$ in [~\citenum{cs}], Fig. 4, where the calculation was done for Cs. The value for Hg is about 1/6 of that for Cs, due to the shorter wavelength of the allowed electric dipole moment transition of Hg, 180 nm compared to 895 nm and 854 nm (D1 and D2 transitions) for Cs, together with the reduced dielectric response of SiO$_2$ at shorter wavelengths, leading to a reduction in the overlap integral. 

The number of bound states is approximately the number of nodes for the least tightly bound WKB solution, which is $N\approx 60 $ for $m=199$, and the two lowest states (via the Schr\"odinger equation) are approximately -0.18 $e$V and -0.21 $e$V.  

Therefore, Eq. (3) of [\citenum{romalis}] needs to be modified, and one possibility is to instead take the average sticking time as
\begin{equation}
\tau_s\propto\sum_{n=1}^N \tau_{0n}e^{|E_{an}|/kT}
\end{equation}
where we allow $\tau_0$ to depend on $n$, and $E_{an}$ are the bound state energies.  
The energy levels are closely spaced, so the possible assumption that only the most tightly bound level contributes significantly to the sum is not appropriate. As an example, assume $\tau_{0n}$ is constant and take the sum of the 10 lowest bound values.  The effective surface energy, implied by the relative rate of change of the sum with temperature, leads to an apparent $E_a\approx -0.16\ e$ V, which is in concordance with the value experimentally determined in [~\citenum{romalis}] using their Eq. (3), and are thought to be abnormally large.  The analysis presented here shows that this large value is expected. In some sense, this measurement provides a means to determine $d_0$ in Eq. (\ref{vdw}) above.

The Hg-CH$_4$ van der Waals binding energy of -0.02 $e$V quoted in [~\citenum{romalis}] is largely irrelevant to this problem because the atom-surface interaction is better understood as a short range van der Walls potential between a dipole and its image in the dielectric surface.\cite{vdw95}

Second, because the average distance from the surface depends on $n$, spin perturbations due to surface interactions will differ between the $n$ states and the correlation times will be affected. In
[~\citenum{romalis}], it is noted that the correlation times appear to be longer than the average sticking times. It is beyond the scope of this short note to analyze the full ramifications of the panoply of surface bound states, however, a reasoned conclusion is that a full analysis requires their inclusion and likely leads to the observed anomalous behavior.

Perhaps the most important consequence of the large number of relatively deep surface states (compared to $k_bT=0.030$ eV) is in regard to the assumption that a precise volume magnetic field average is provided by a \hg magnetometer cell.  For an extremely precise magnetometry system, as has been proposed, the surface dwell time (during which most spin polarization survives) will need to be more carefully considered, as the dwell time can lead to corrections, for example, for localized magnetic fields that present a relatively large gradient.

This work was supported by Yale University.

\bibliography{lamoreaux_rev_2}%

\begin{thebibliography}{7}%
\makeatletter
\providecommand \@ifxundefined [1]{%
 \@ifx{#1\undefined}
}%
\providecommand \@ifnum [1]{%
 \ifnum #1\expandafter \@firstoftwo
 \else \expandafter \@secondoftwo
 \fi
}%
\providecommand \@ifx [1]{%
 \ifx #1\expandafter \@firstoftwo
 \else \expandafter \@secondoftwo
 \fi
}%
\providecommand \natexlab [1]{#1}%
\providecommand \enquote  [1]{``#1''}%
\providecommand \bibnamefont  [1]{#1}%
\providecommand \bibfnamefont [1]{#1}%
\providecommand \citenamefont [1]{#1}%
\providecommand \href@noop [0]{\@secondoftwo}%
\providecommand \href [0]{\begingroup \@sanitize@url \@href}%
\providecommand \@href[1]{\@@startlink{#1}\@@href}%
\providecommand \@@href[1]{\endgroup#1\@@endlink}%
\providecommand \@sanitize@url [0]{\catcode `\\12\catcode `\$12\catcode
  `\&12\catcode `\#12\catcode `\^12\catcode `\_12\catcode `\%12\relax}%
\providecommand \@@startlink[1]{}%
\providecommand \@@endlink[0]{}%
\providecommand \url  [0]{\begingroup\@sanitize@url \@url }%
\providecommand \@url [1]{\endgroup\@href {#1}{\urlprefix }}%
\providecommand \urlprefix  [0]{URL }%
\providecommand \Eprint [0]{\href }%
\providecommand \doibase [0]{http://dx.doi.org/}%
\providecommand \selectlanguage [0]{\@gobble}%
\providecommand \bibinfo  [0]{\@secondoftwo}%
\providecommand \bibfield  [0]{\@secondoftwo}%
\providecommand \translation [1]{[#1]}%
\providecommand \BibitemOpen [0]{}%
\providecommand \bibitemStop [0]{}%
\providecommand \bibitemNoStop [0]{.\EOS\space}%
\providecommand \EOS [0]{\spacefactor3000\relax}%
\providecommand \BibitemShut  [1]{\csname bibitem#1\endcsname}%
\let\auto@bib@innerbib\@empty
\bibitem [{\citenamefont {Ito}\ \emph {et~al.}(2018)\citenamefont {Ito} \emph
  {et~al.}}]{lanledm}%
  \BibitemOpen
  \bibfield  {author} {\bibinfo {author} {\bibfnamefont {T.~M.}\ \bibnamefont
  {Ito}} \emph {et~al.},\ }\bibfield  {title} {\enquote {\bibinfo {title}
  {Performance of the upgraded ultracold neutron source at los alamos national
  laboratory and its implication for a possible neutron electric dipole moment
  experiment},}\ }\href {\doibase 10.1103/PhysRevC.97.012501} {\bibfield
  {journal} {\bibinfo  {journal} {Phys. Rev. C}\ }\textbf {\bibinfo {volume}
  {97}},\ \bibinfo {pages} {012501} (\bibinfo {year} {2018})}\BibitemShut
  {NoStop}%
\bibitem [{\citenamefont {Ayres}\ \emph {et~al.}(2021)\citenamefont {Ayres}
  \emph {et~al.}}]{psiedm}%
  \BibitemOpen
  \bibfield  {author} {\bibinfo {author} {\bibfnamefont {N.~J.}\ \bibnamefont
  {Ayres}} \emph {et~al.},\ }\bibfield  {title} {\enquote {\bibinfo {title}
  {The design of the n2edm experiment},}\ }\href {\doibase
  10.1140/epjc/s10052-021-09298-z} {\bibfield  {journal} {\bibinfo  {journal}
  {The European Physical Journal C}\ }\textbf {\bibinfo {volume} {81}},\
  \bibinfo {pages} {512} (\bibinfo {year} {2021})}\BibitemShut {NoStop}%
\bibitem [{\citenamefont {Romalis}\ and\ \citenamefont {Lin}(2004)}]{romalis}%
  \BibitemOpen
  \bibfield  {author} {\bibinfo {author} {\bibfnamefont {M.~V.}\ \bibnamefont
  {Romalis}}\ and\ \bibinfo {author} {\bibfnamefont {L.}~\bibnamefont {Lin}},\
  }\bibfield  {title} {\enquote {\bibinfo {title} {{Surface nuclear spin
  relaxation of $^{199}$Hg}},}\ }\href@noop {} {\bibfield  {journal} {\bibinfo
  {journal} {The Journal of Chemical Physics}\ }\textbf {\bibinfo {volume}
  {120}},\ \bibinfo {pages} {1511--1515} (\bibinfo {year} {2004})}\BibitemShut
  {NoStop}%
\bibitem [{\citenamefont {{J.M. Pendlebury (deceased)}}(1987)}]{jmp}%
  \BibitemOpen
  \bibfield  {author} {\bibinfo {author} {\bibnamefont {{J.M. Pendlebury
  (deceased)}}},\ }\href@noop {} {}\bibinfo {howpublished} {Private
  Communication} (\bibinfo {year} {1987})\BibitemShut {NoStop}%
\bibitem [{\citenamefont {Driehuys}, \citenamefont {Cates},\ and\ \citenamefont
  {Happer}(1995)}]{happer}%
  \BibitemOpen
  \bibfield  {author} {\bibinfo {author} {\bibfnamefont {B.}~\bibnamefont
  {Driehuys}}, \bibinfo {author} {\bibfnamefont {G.~D.}\ \bibnamefont {Cates}},
  \ and\ \bibinfo {author} {\bibfnamefont {W.}~\bibnamefont {Happer}},\
  }\bibfield  {title} {\enquote {\bibinfo {title} {Surface relaxation
  mechanisms of laser-polarized ${}^{129}\mathrm{Xe}$},}\ }\href {\doibase
  10.1103/PhysRevLett.74.4943} {\bibfield  {journal} {\bibinfo  {journal}
  {Phys. Rev. Lett.}\ }\textbf {\bibinfo {volume} {74}},\ \bibinfo {pages}
  {4943--4946} (\bibinfo {year} {1995})}\BibitemShut {NoStop}%
\bibitem [{\citenamefont {Fichet}\ \emph {et~al.}(1995)\citenamefont {Fichet},
  \citenamefont {Schuller}, \citenamefont {Bloch},\ and\ \citenamefont
  {Ducloy}}]{vdw95}%
  \BibitemOpen
  \bibfield  {author} {\bibinfo {author} {\bibfnamefont {M.}~\bibnamefont
  {Fichet}}, \bibinfo {author} {\bibfnamefont {F.}~\bibnamefont {Schuller}},
  \bibinfo {author} {\bibfnamefont {D.}~\bibnamefont {Bloch}}, \ and\ \bibinfo
  {author} {\bibfnamefont {M.}~\bibnamefont {Ducloy}},\ }\bibfield  {title}
  {\enquote {\bibinfo {title} {van der waals interactions between excited-state
  atoms and dispersive dielectric surfaces},}\ }\href {\doibase
  10.1103/PhysRevA.51.1553} {\bibfield  {journal} {\bibinfo  {journal} {Phys.
  Rev. A}\ }\textbf {\bibinfo {volume} {51}},\ \bibinfo {pages} {1553--1564}
  (\bibinfo {year} {1995})}\BibitemShut {NoStop}%
\bibitem [{\citenamefont {Stern}, \citenamefont {Alton},\ and\ \citenamefont
  {Kimble}(2011)}]{cs}%
  \BibitemOpen
  \bibfield  {author} {\bibinfo {author} {\bibfnamefont {N.~P.}\ \bibnamefont
  {Stern}}, \bibinfo {author} {\bibfnamefont {D.~J.}\ \bibnamefont {Alton}}, \
  and\ \bibinfo {author} {\bibfnamefont {H.~J.}\ \bibnamefont {Kimble}},\
  }\bibfield  {title} {\enquote {\bibinfo {title} {Simulations of atomic
  trajectories near a dielectric surface},}\ }\href {\doibase
  10.1088/1367-2630/13/8/085004} {\bibfield  {journal} {\bibinfo  {journal}
  {New Journal of Physics}\ }\textbf {\bibinfo {volume} {13}},\ \bibinfo
  {pages} {085004} (\bibinfo {year} {2011})}\BibitemShut {NoStop}%
\end{thebibliography}%

\end{document}